\documentclass[runningheads]{llncs}

\usepackage{amsmath}
\usepackage{amssymb}
\usepackage{color}
\usepackage{graphicx}
\usepackage{hyperref}

\usepackage{multirow}
\usepackage{subfig}
\usepackage{threeparttable}
\usepackage{xspace}

\newcommand{\ise}{\mathtt{isE}\xspace}
\newcommand{\isc}{\mathtt{isC}\xspace}
\newcommand{\isl}{\mathtt{isL}\xspace}
\newcommand{\localfreq}{\mathtt{lf}\xspace}
\newcommand{\vfreq}{\mathtt{vf}\xspace}
\newcommand{\selfinfo}{\mathtt{si}\xspace}
\newcommand{\globalfreq}{\mathtt{gf}\xspace}

\begin{document}

\title{ESBM: An Entity Summarization BenchMark}

\author{Qingxia Liu\inst{1} \and
Gong Cheng\inst{1} \and
Kalpa Gunaratna\inst{2} \and
Yuzhong Qu\inst{1}}
\authorrunning{Q. Liu et al.}

\institute{National Key Laboratory for Novel Software Technology, Nanjing University, China\\
\email{qxliu2013@smail.nju.edu.cn, \{gcheng,yzqu\}@nju.edu.cn}
\and
Samsung Research America, Mountain View CA, USA\\
\email{k.gunaratna@samsung.com}}

\maketitle

\begin{abstract}

Entity summarization is the problem of computing an optimal compact summary for an entity by selecting a size-constrained subset of triples from RDF data. Entity summarization supports a multiplicity of applications and has led to fruitful research. However, there is a lack of evaluation efforts that cover the broad spectrum of existing systems. One reason is a lack of benchmarks for evaluation. Some benchmarks are no longer available, while others are small and have limitations. In this paper, we create an Entity Summarization BenchMark (ESBM) which overcomes the limitations of existing benchmarks and meets standard desiderata for a benchmark. Using this largest available benchmark for evaluating general-purpose entity summarizers, we perform the most extensive experiment to date where 9~existing systems are compared. Considering that all of these systems are unsupervised, we also implement and evaluate a supervised learning based system for reference.


\end{abstract}

\section{Introduction}

\begin{figure}[!t]
    \centering
    \includegraphics[width=0.85\linewidth]{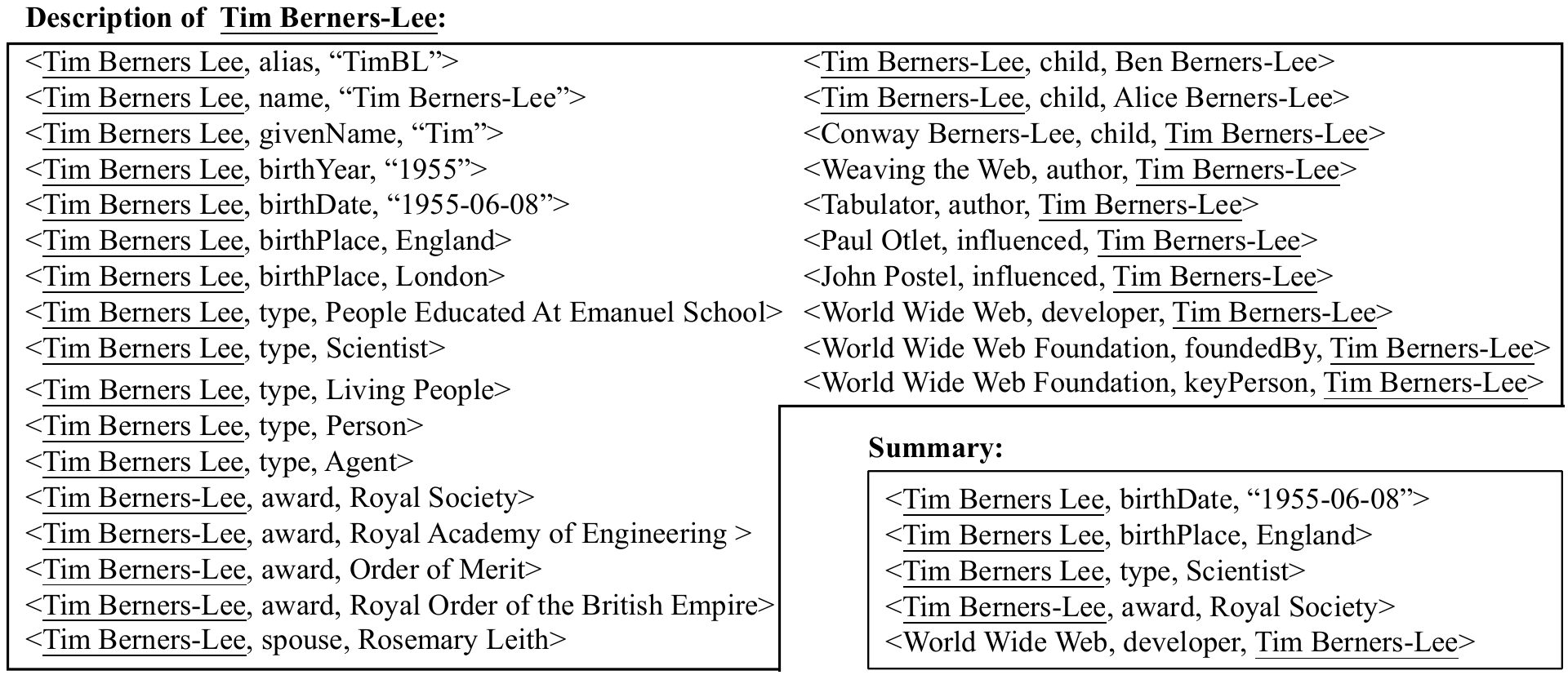}
    \caption{Description of entity \texttt{Tim Berners-Lee} and a summary thereof.}
    \label{fig:example}
\end{figure}

RDF data describes entities with triples representing property values. In an RDF dataset, the description of an entity comprises all the RDF triples where the entity appears as the subject or the object. An example entity description is shown in Fig.~\ref{fig:example}. Entity descriptions can be large. An entity may be described in dozens or hundreds of triples, exceeding the capacity of a typical user interface. A user served with all of those triples may suffer information overload and find it difficult to quickly identify the small set of triples that are truly needed. To solve the problem, an established research topic is \emph{entity summarization}~\cite{survey}, which aims to compute an optimal compact summary for the entity by selecting a size-constrained subset of triples. An example entity summary under the size constraint of 5~triples is shown in the bottom right corner of Fig.~\ref{fig:example}.

Entity summarization supports a multiplicity of applications~\cite{thesiskalpa,thesisandreas}. Entity summaries constitute entity cards displayed in search engines~\cite{dynes}, provide background knowledge for enriching documents~\cite{trank}, and facilitate research activities with humans in the loop~\cite{entityresolution,entitylinking}. This far-reaching application has led to fruitful research as reviewed in our recent survey paper~\cite{survey}. Many entity summarizers have been developed, most of which generate summaries for general purposes.

\textbf{Research Challenges.}
However, two challenges face the research community. First, there is a \emph{lack of benchmarks} for evaluating entity summarizers. As shown in Table~\ref{tab:benchmarks}, some benchmarks are no longer available. Others are available~\cite{game,faces,facese} but they are small and have limitations. Specifically, \cite{game}~has a task-specific nature, and \cite{faces,facese}~exclude classes and/or literals. These benchmarks could not support a comprehensive evaluation of general-purpose entity summarizers. Second, there is a \emph{lack of evaluation efforts} that cover the broad spectrum of existing systems to compare their performance and assist practitioners in choosing solutions appropriate to their applications.

\textbf{Contributions.}
We address the challenges with two contributions. First, we create an \textbf{E}ntity \textbf{S}ummarization \textbf{B}ench\textbf{M}ark (ESBM) which overcomes the limitations of existing benchmarks and meets the desiderata for a successful benchmark~\cite{benchmark}. ESBM has been published on GitHub with extended documentation and a permanent identifier on \texttt{w3id.org}\footnote{\url{https://w3id.org/esbm}} under the ODC-By license. As the largest available benchmark for evaluating general-purpose entity summarizers, ESBM contains 175~heterogeneous entities sampled from two datasets, for which 30~human experts create 2,100~general-purpose ground-truth summaries under two size constraints. Second, using ESBM, we evaluate 9~existing general-purpose entity summarizers. It represents the most extensive evaluation effort to date. Considering that existing systems are unsupervised, we also implement and evaluate a supervised learning based entity summarizer for reference.

In this paper, for the first time we comprehensively describe the creation and use of ESBM. We report ESBM~v1.2---the latest version, while early versions have successfully supported the entity summarization shared task at the EYRE~2018 workshop\footnote{\url{https://sites.google.com/view/eyre18/sharedtasks}} and the EYRE~2019 workshop.\footnote{\url{https://sites.google.com/view/eyre19/sharedtasks}} We will also educate on the use of ESBM at an ESWC~2020 tutorial on entity summarization\footnote{\url{https://sites.google.com/view/entity-summarization-tutorials/eswc2020}}.

The remainder of the paper is organized as follows. Section~\ref{sec:related-work} reviews related work and limitations of existing benchmarks. Section~\ref{sec:creating} describes the creation of ESBM, which is analyzed in Section~\ref{sec:analyzing}. Section~\ref{sec:evaluating} presents our evaluation. In Section~\ref{sec:discussion-future-work} we discuss limitations of our study and perspectives for future work.

\begin{table}[!t]
	\caption{Existing benchmarks for evaluating entity summarization.}
	\label{tab:benchmarks}
	\centering
	\begin{tabular}{|l|l|r|l|}
		\hline
		 & Dataset & Number of entities & Availability \\
		\hline
		WhoKnows?Movies!~\cite{game} & Freebase & 60 & \textbf{Available}\footnotemark[1] \\
		Langer et al.~\cite{assign} & DBpedia & 14 & Unavailable \\
		FRanCo~\cite{franco} & DBpedia & 265 & Unavailable \\
		Benchmark for evaluating RELIN~\cite{relin} & DBpedia & 149 & Unavailable \\
		Benchmark for evaluating DIVERSUM~\cite{diversum} & IMDb & 20 & Unavailable \\
		Benchmark for evaluating FACES~\cite{faces} & DBpedia & 50 & \textbf{Available}\footnotemark[2] \\
		Benchmark for evaluating FACES-E~\cite{facese} & DBpedia & 80 & \textbf{Available}\footnotemark[2] \\
		\hline
	\end{tabular}
\end{table}
\footnotetext[1]{\url{http://yovisto.com/labs/iswc2012}}
\footnotetext[2]{\url{http://wiki.knoesis.org/index.php/FACES}}

\section{Related Work}
\label{sec:related-work}

We review methods and evaluation efforts for entity summarization.

\textbf{Methods for Entity Summarization.}
In a recent survey~\cite{survey} we have categorized the broad spectrum of research on entity summarization. Below we briefly review \emph{general-purpose} entity summarizers which mainly rely on generic technical features that can apply to a wide range of domains and applications. We will not address methods that are domain-specific (e.g.,~for movies~\cite{movie} or timelines~\cite{timeline}), task-specific (e.g.,~for facilitating entity resolution~\cite{entityresolution} or entity linking~\cite{entitylinking}), or context-aware (e.g.,~contextualized by a document~\cite{trank} or a query~\cite{dynes}).

RELIN~\cite{relin} uses a weighted PageRank model to rank triples according to their statistical informativeness and relatedness. DIVERSUM~\cite{diversum} ranks triples by property frequency and generates a summary with a strong constraint that avoids selecting triples having the same property. SUMMARUM~\cite{summarum} and LinkSUM~\cite{linksum} mainly rank triples by the PageRank scores of property values that are entities. LinkSUM also considers backlinks from values. FACES~\cite{faces}, and its extension FACES-E~\cite{facese} which adds support for literals, cluster triples by their bag-of-words based similarity and choose top-ranked triples from as many different clusters as possible. Triples are ranked by statistical informativeness and property value frequency. CD~\cite{cd} models entity summarization as a quadratic knapsack problem that maximizes the statistical informativeness of the selected triples and in the meantime minimizes the string, numerical, and logical similarity between them.
In ES-LDA~\cite{eslda}, ES-LDA$_{ext}$~\cite{esldaext}, and MPSUM~\cite{mpsum}, a Latent Dirichlet Allocation (LDA) model is learned where properties are treated as topics, and each property is a distribution over all the property values. Triples are ranked by the probabilities of properties and values. MPSUM further avoids selecting triples having the same property. BAFREC~\cite{bafrec} categorizes triples into meta-level and data-level. It ranks meta-level triples by their depths in an ontology and ranks data-level triples by property and value frequency. Triples having textually similar properties are penalized to improve diversity. KAFCA~\cite{kafca} ranks triples by the depths of properties and values in a hierarchy constructed by performing the Formal Concept Analysis (FCA). It tends to select triples containing infrequent properties but frequent values, where frequency is computed at the word level.

\textbf{Limitations of Existing Benchmarks.}
For evaluating entity summarization, compared with task completion based \emph{extrinsic evaluation}, ground truth based \emph{intrinsic evaluation} is more popular because it is easy to perform and the results are reproducible. Its idea is to create a benchmark consisting of human-made ground-truth summaries, and then compute how much a machine-generated summary is close to a ground-truth summary.

Table~\ref{tab:benchmarks} lists known benchmarks, including dedicated benchmarks~\cite{game,assign,franco} and those created for evaluating a particular entity summarizer~\cite{relin,diversum,faces,facese}. It is not surprising that these benchmarks are not very large since it is expensive to manually create high-quality summaries for a large set of entities. Unfortunately, some of these benchmarks are not publicly available at this moment. Three are available~\cite{game,faces,facese} but they are relatively small and have limitations. Specifically, WhoKnows?Movies!~\cite{game} is not a set of ground-truth summaries but annotates each triple with the ratio of movie questions that were correctly answered based on that triple, as an indicator of its importance. This kind of task-specific ground truth may not be suitable for evaluating general-purpose entity summarizers. The other two available benchmarks were created for evaluating FACES/-E~\cite{faces,facese}. Classes and/or literals are not included because they could not be processed by FACES/-E and hence were filtered out. Such benchmarks could not comprehensively evaluate most of the existing entity summarizers~\cite{relin,diversum,cd,mpsum,bafrec,kafca} that can handle classes and literals. These limitations of available benchmarks motivated us to create a new ground truth consisting of \emph{general-purpose summaries} for a \emph{larger set of entities} involving \emph{more comprehensive triples} where property values can be entities, classes, or literals.
\section{Creating ESBM}
\label{sec:creating}

To overcome the above-mentioned limitations of existing benchmarks, we created a new benchmark called ESBM. To date, it is the largest available benchmark for evaluating general-purpose entity summarizers. In this section, we will first specify our design goals. Then we describe the selection of entity descriptions and the creation of ground-truth summaries. We partition the data to support cross-validation for parameter fitting. Finally we summarize how our design goals are achieved and how ESBM meets standard desiderata for a benchmark.

\subsection{Design Goals}
\label{sec:creating-goal}

The creation of ESBM has two main design goals. First, a successful benchmark should meet seven desiderata~\cite{benchmark}: accessibility, affordability, clarity, relevance, solvability, portability, and scalability, which we will detail in Section~\ref{sec:creating-conclusion}. Our design of ESBM aims to satisfy these basic requirements. Second, in Section~\ref{sec:related-work} we discussed the limitations of available benchmarks, including task specificness, small size, and triple incomprehensiveness. Besides, all the existing benchmarks use a single dataset and hence may weaken the generalizability of evaluation results. We aim to overcome these limitations when creating ESBM. In Section~\ref{sec:creating-conclusion} we will summarize how our design goals are achieved.

\subsection{Entity Descriptions}

To choose entity descriptions to summarize, we sample entities from selected datasets and filter their triples. The process is detailed below.

\textbf{Datasets.}
We sample entities from two datasets of different kinds: an encyclopedic dataset and a domain-specific dataset. For the encyclopedic dataset we choose DBpedia~\cite{dbpedia}, which has been used in other benchmarks~\cite{assign,franco,relin,faces,facese}. We use the English version of DBpedia 2015-10\footnote{\url{http://wiki.dbpedia.org/dbpedia-dataset-version-2015-10}}---the latest version when we started to create ESBM. For the domain-specific dataset we choose LinkedMDB~\cite{lmdb}, which is a popular movie database. The movie domain is also the focus of some existing benchmarks~\cite{game,diversum} possibly because this domain is familiar to the lay audience so that it would be easy to find qualified human experts to create ground-truth summaries. We use the latest available version of LinkedMDB.\footnote{\url{http://www.cs.toronto.edu/~oktie/linkedmdb/linkedmdb-latest-dump.zip}}

\textbf{Entities.}
For DBpedia we sample entities from five large classes: \texttt{Agent}, \texttt{Event}, \texttt{Location}, \texttt{Species}, and \texttt{Work}. They collectively contain 3,501,366~entities (60\%) in the dataset. For LinkedMDB we sample from  \texttt{Film} and \texttt{Person}, which contain 159,957~entities (24\%) in the dataset. Entities from different classes are described by very different properties as we will see in Section~\ref{sec:enthete}, and hence help to assess the generalizability of an entity summarizer. According to the human efforts we could afford, from each class we randomly sample 25~entities. The total number of selected entities is~175. Each selected entity should be described in at least 20~triples so that summarization would not be a trivial task. This requirement follows common practice in the literature~\cite{franco,relin,diversum,faces} where a minimum constraint in the range of 10--20 was posed.

\begin{figure*}[!t]
	\centering
	\subfloat[Average number of triples describing an entity.]
	{
		\includegraphics[width=0.47\linewidth]{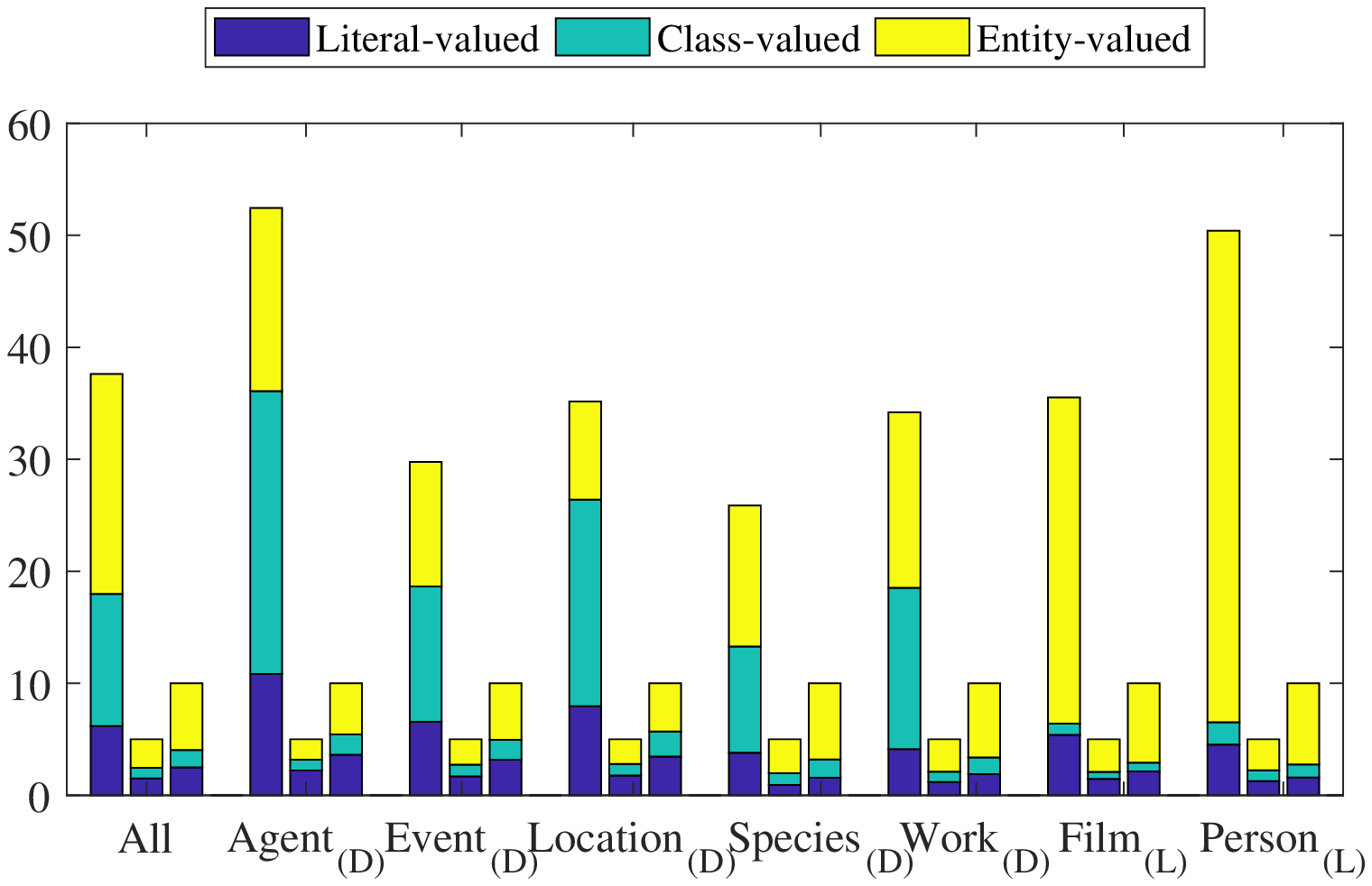}
		\label{fig:composition-triple}
	}
	\hfill
	\subfloat[Average number of distinct properties describing an entity.]
	{
	    \includegraphics[width=0.47\linewidth]{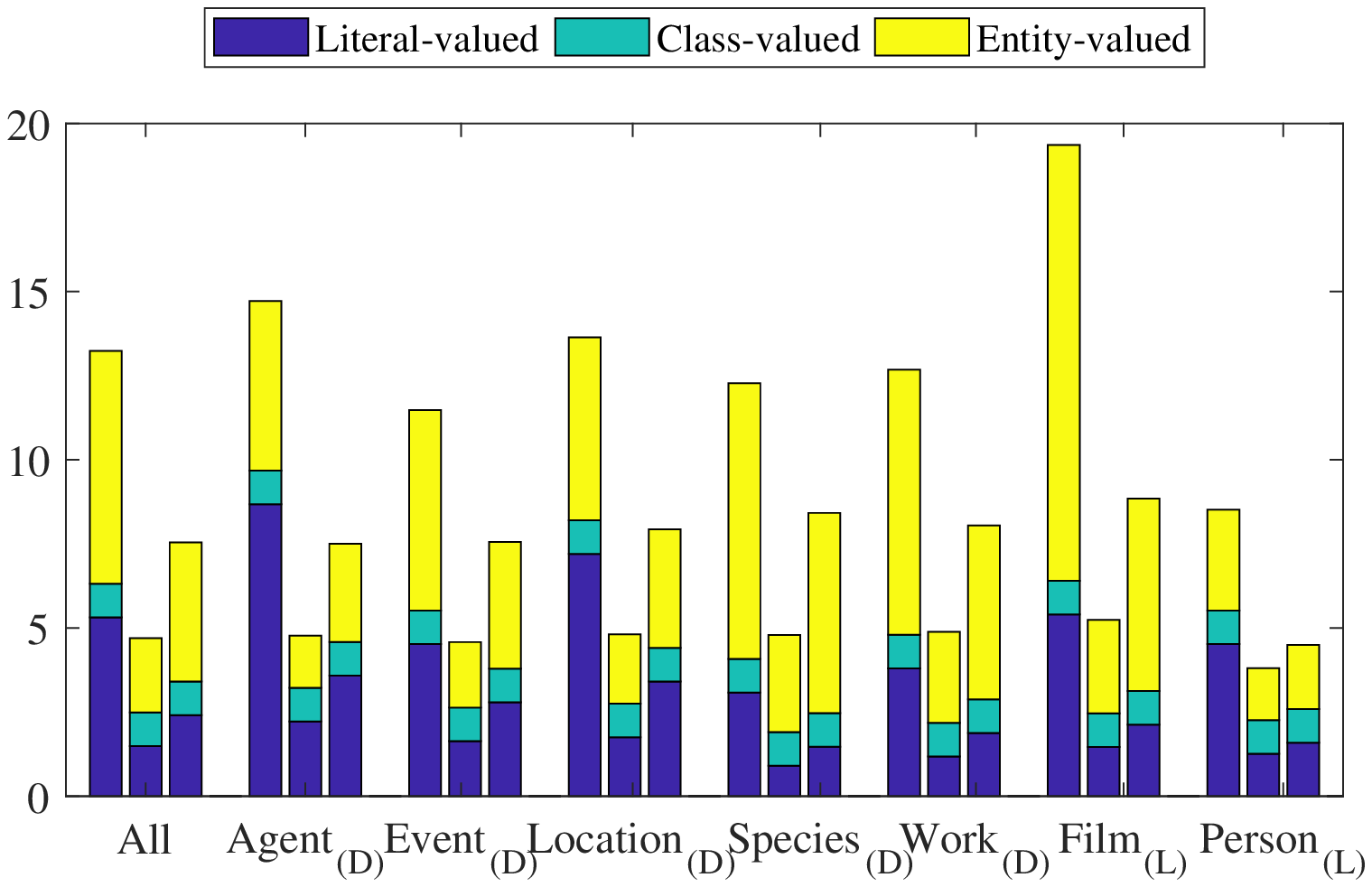}
	    \label{fig:composition-property}
	}
	\caption{Composition of entity descriptions (the left bar in each group), top-5 ground-truth summaries (the middle bar), and top-10 ground-truth summaries (the right bar), grouped by class in DBpedia~(D) and LinkedMDB~(L).}
	\label{fig:composition}
\end{figure*}

\textbf{Triples.}
For DBpedia, entity descriptions comprise triples in the following dump files: \emph{instance types}, \emph{instance types transitive}, \emph{YAGO types}, \emph{mappingbased literals}, \emph{mappingbased objects}, \emph{labels}, \emph{images}, \emph{homepages}, \emph{persondata}, \emph{geo coordinates mappingbased}, and \emph{article categories}. We do not import dump files that provide metadata about Wikipedia articles such as \emph{page links} and \emph{page length}. We do not import \emph{short abstracts} and \emph{long abstracts} as they provide handcrafted textual entity summaries; it would be inappropriate to include them in a benchmark for evaluating entity summarization. For LinkedMDB we import all the triples in the dump file except \texttt{sameAs} links which do not express facts about entities but are of more technical nature. Finally, as shown in Fig.~\ref{fig:composition-triple} (the left bar in each group), the mean number of triples in an entity description is in the range of 25.88--52.44 depending on the class, and the overall mean value is~37.62.

\subsection{Ground-Truth Summaries}

We invite 30~researchers and students to create ground-truth summaries for entity descriptions. All the participants are familiar with RDF.

\textbf{Task Assignment.}
Each participant is assigned 35~entities consisting of 5~entities randomly selected from each of the 7~classes in ESBM. The assignment is controlled to ensure that each entity in ESBM is processed by 6~participants. A participant creates two summaries for each entity description by selecting different numbers of triples: a \emph{top-5 summary} containing 5~triples, and a \emph{top-10 summary} containing 10~triples. Therefore, we will be able to evaluate entity summarizers under different size constraints. The choice of these two numbers follows previous work~\cite{relin,faces,facese}. Participants work independently and they may create different summaries for an entity. It is not feasible to ask participants to reach an agreement. It is also not reasonable to merge different summaries into a single version. So we keep different summaries and will use all of them in the evaluation. The total number of ground-truth summaries is $175 \cdot 6 \cdot 2 = 2,100$.

\begin{figure}[!t]
\includegraphics[width=\textwidth]{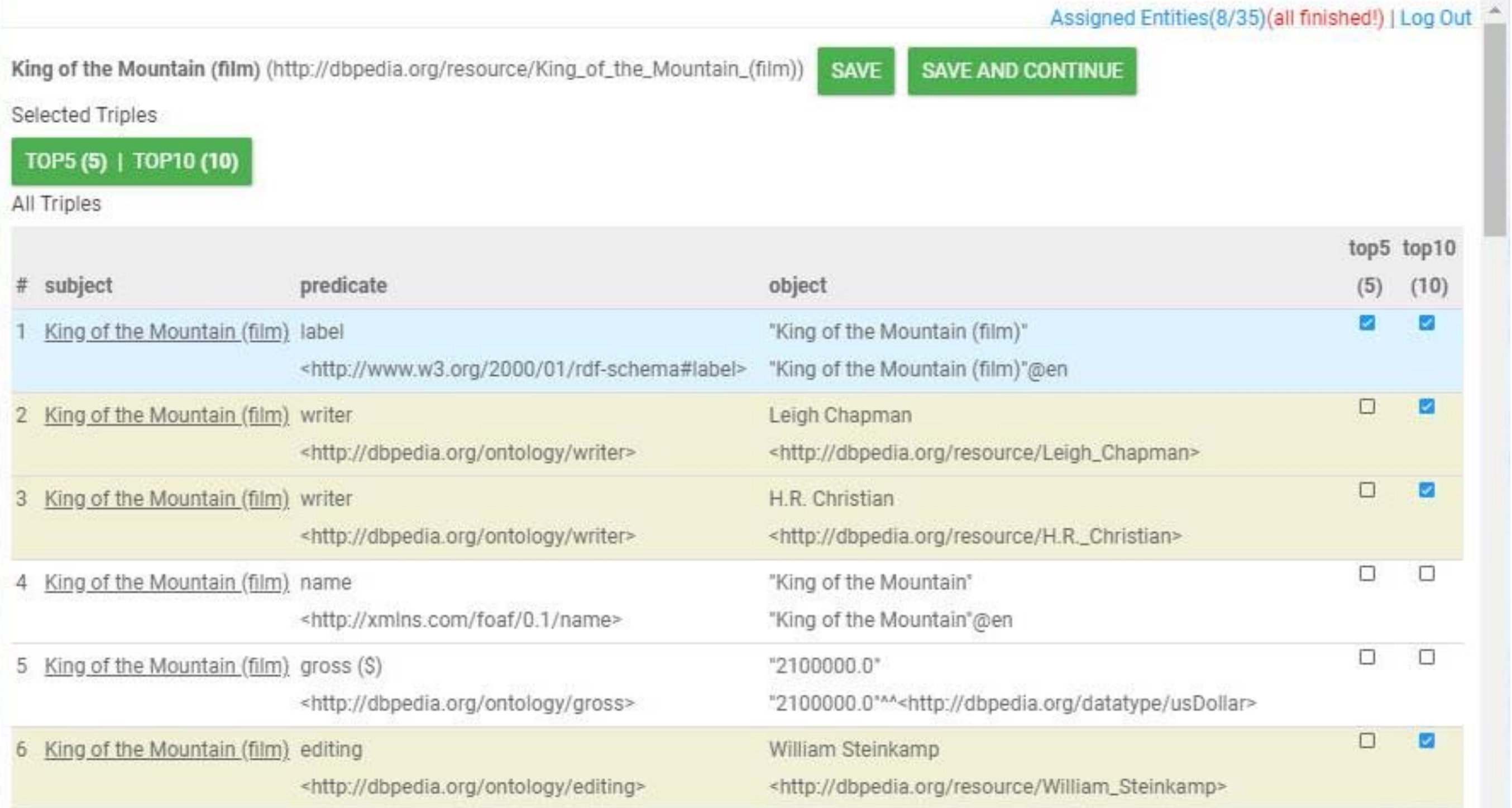}
\caption{User interface for creating ground-truth entity summaries. }
\label{fig:screen}
\end{figure}

\textbf{Procedure.}
Participants are instructed to create \emph{general-purpose summaries} that are not specifically created for any particular task. They read and select triples using a Web-based user interface shown in Fig.~\ref{fig:screen}. All the triples in an entity description are listed in random order but those having a common property are placed together for convenient reading and comparison. For IRIs, their human-readable labels (\texttt{rdfs:label}) are shown if available. To help participants understand a property value that is an unfamiliar entity, a click on it will open a pop-up showing a short textual description extracted from the first paragraph of its Wikipedia/IMDb page. Any triple can be selected into the top-5 summary, the top-10 summary, or both. The top-5 summary is not required to be a subset of the top-10 summary.

\subsection{Training, Validation, and Test Sets}
\label{sec:creating-train-test}

Some entity summarizers need to tune hyperparameters or fit models. To make their evaluation results comparable with each other, we specify a split of our data into training, validation, and test sets. We provide a partition of the 175~entities in ESBM into 5~equally sized subsets $P_0, \ldots, P_4$ to support 5-fold cross-validation. Entities of each class are partitioned evenly among the subsets. For $0 \leq i \leq 4$, the $i$-th fold uses $P_i,P_{i+1 \text{ mod } 5},P_{i+2 \text{ mod } 5}$ as the training set (e.g.,~for model fitting), uses~$P_{i+3 \text{ mod } 5}$ for validation (e.g.,~tuning hyperparameters), and retains~$P_{i+4 \text{ mod } 5}$ as the test set. Evaluation results are averaged over the 5~folds.

\subsection{Conclusion}
\label{sec:creating-conclusion}

ESBM overcomes the limitations of available benchmarks discussed in Section~\ref{sec:related-work}. It contains 175~entities which is 2--3~times as large as available benchmarks~\cite{game,faces,facese}. In ESBM, property values are not filtered as in~\cite{faces,facese} but can be any entity, class, or literal. Different from the task-specific nature of~\cite{game}, ESBM provides general-purpose ground-truth summaries for evaluating general-purpose entity summarizers.

Besides, ESBM meets the seven desiderata proposed in~\cite{benchmark} as follows.
\begin{itemize}
    \item \textbf{Accessibility.} ESBM is publicly available and has a permanent identifier on \texttt{w3id.org}.
    \item \textbf{Affordability.} ESBM is with an open-source program and example code for evaluation. The cost of using ESBM is minimized.
    \item \textbf{Clarity.} ESBM is documented clearly and concisely.
    \item \textbf{Relevance.} ESBM samples entities from two real datasets that have been widely used. The summarization tasks are natural and representative.
    \item \textbf{Solvability.} An entity description in ESBM has at least 20~triples and a mean number of 37.62~triples, from which 5~or 10~triples are to be selected. The summarization tasks are not trivial and not too difficult.
    \item \textbf{Portability.} ESBM can be used to evaluate any general-purpose entity summarizer that can process RDF data.
    \item \textbf{Scalability.} ESBM samples 175~entities from 7~classes. It is reasonably large and diverse to evaluate mature entity summarizers but is not too large to evaluate research prototypes.
\end{itemize}

However, ESBM has its own limitations, which we will discuss in Section~\ref{sec:discussion-future-work}.
\section{Analyzing ESBM}
\label{sec:analyzing}

In this section, we will first characterize ESBM by providing some basic statistics and analyzing the triple composition and heterogeneity of entity descriptions. Then we compute inter-rater agreement to show how much consensus exists in the ground-truth summaries given by different participants.

\subsection{Basic Statistics}

The 175~entity descriptions in ESBM collectively contain 6,584~triples, of which 37.44\%~are selected into at least one top-5 summary and 58.15\%~appear in at least one top-10 summary, showing a wide selection by the participants. However, many of them are selected only by a single participant; 20.46\% 
and 40.23\% 
are selected by different participants into top-5 and top-10 summaries, respectively. We will further analyze inter-rater agreement in Section~\ref{sec:analyzing-agreement}.

We calculate the overlap between the top-5 and the top-10 summaries created by the same participant for the same entity. The mean overlap is in the range of 4.80--4.99~triples depending on the class, and the overall mean value is~4.91, showing that the top-5 summary is usually a subset of the top-10 summary.

\subsection{Triple Composition}\label{sec:tcomp}

In Fig.~\ref{fig:composition} we present the composition of entity descriptions (the left bar in each group) and their ground-truth summaries (the middle bar for top-5 and the right bar for top-10) in ESBM, in terms of the average number of triples describing an entity (Fig.~\ref{fig:composition-triple}) and in terms of the average number of distinct properties describing an entity (Fig.~\ref{fig:composition-property}). Properties are divided into literal-valued, class-valued, and entity-valued. Triples are divided accordingly.

In Fig.~\ref{fig:composition-triple}, both class-valued and entity-valued triples occupy a considerable proportion of the entity descriptions in DBpedia. Entity-valued triples predominate in LinkedMDB. Literal-valued triples account for a small proportion in both datasets. However, they constitute 30\% in top-5 ground-truth summaries and 25\% in top-10 summaries. Entity summarizers that cannot process literals~\cite{summarum,linksum,faces,eslda} have to ignore these notable proportions, thereby significantly influencing their performance.

In Fig.~\ref{fig:composition-property}, in terms of distinct properties, entity-valued and literal-valued triples have comparable numbers in entity descriptions since many entity-valued properties are multi-valued. Specifically, an entity is described by 13.24~distinct properties, including 5.31~literal-valued (40\%) and 6.93~entity-valued (52\%). Multi-valued properties appear in every entity description and they constitute 35\% of the triples. However, in top-5 ground-truth summaries, the average number of distinct properties is~4.70 and is very close to~5, indicating that the participants are not inclined to select multiple values of a property. Entity summarizers that prefer diverse properties~\cite{diversum,faces,facese,cd,mpsum,bafrec} may exhibit good performance.

\begin{figure}[!t]
	\centering
	\includegraphics[width=0.75\linewidth]{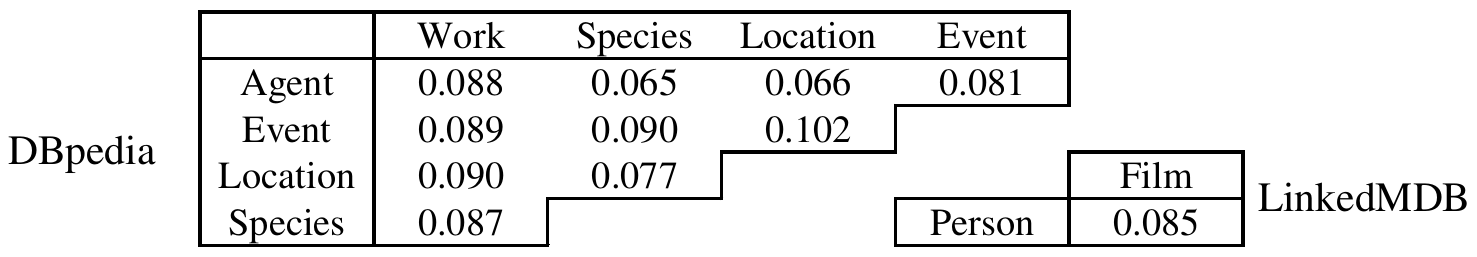}
	\caption{Jaccard similarity between property sets describing different classes.}
	\label{fig:prop_overlap}
\end{figure}

\begin{table}[!t]
\caption{Popular properties in ground-truth summaries.}
\label{tab:popular-properties}
\resizebox{\columnwidth}{!}{
\begin{tabular}{|l|l|l|l|l|l|l|l|l|l|l|l|l|l|}
\hline
\multicolumn{7}{|c|}{In top-5 summaries} & \multicolumn{7}{|c|}{In top-10 summaries} \\
\hline
Agent & Event & Location & Species & Work & Film & Person & Agent & Event & Location & Species & Work & Film & Person \\
\hline
type & type & type & type & type & director & type & type & type & type & family & type & director & type \\
birthDate & date & country & family &  & type & actor & subject & subject & country & type & subject & actor & actor \\
 &  &  &  &  &  &  & birthDate & date & subject & order & genre & type & label \\
 &  &  &  &  &  &  &  & label &  & class &  & writer & page \\
 &  &  &  &  &  &  &  &  &  & genus &  & producer &  \\
 &  &  &  &  &  &  &  &  &  & subject &  & date &  \\
 &  &  &  &  &  &  &  &  &  & kingdom &  & language &  \\

\hline
\end{tabular}
}
\end{table}

\subsection{Entity Heterogeneity}\label{sec:enthete}

Entities from different classes are described by different sets of properties. For each class we identify the set of properties describing at least one entity from the class. The Jaccard similarity between properties sets for each pair of classes is very low, as shown in Fig.~\ref{fig:prop_overlap}. Such heterogeneous entity descriptions help to assess the generalizability of an entity summarizer.

Table~\ref{tab:popular-properties} shows popular properties that appear in at least 50\% of the ground-truth summaries for each class.
Some universal properties like \texttt{rdf:type} and \texttt{dct:subject} are popular for most classes. We also see class-specific properties, e.g.,~\texttt{dbo:birthDate} for \texttt{Agent}, \texttt{dbo:family} for \texttt{Species}. However, the results suggest that it would be unrealistic to generate good summaries by manually selecting properties for each class. For example, among 13.24~distinct properties describing an entity, only 1--2~are popular in top-5 ground-truth summaries. The importance of properties is generally contextualized by concrete entities.

\subsection{Inter-Rater Agreement}
\label{sec:analyzing-agreement}

Recall that each entity in ESBM has six top-5 ground-truth summaries and six top-10 summaries created by different participants. We calculate the average overlap between these summaries in terms of the number of common triples they contain. As shown in Table~\ref{tab:overlap}, the results are generally comparable with those reported for other benchmarks in the literature. There is a moderate degree of agreement between the participants.

\begin{table}[!t]
    \centering
	\caption{Inter-rater agreement.}
	\label{tab:overlap}
	\begin{tabular}{|l|cccc|}
		\hline
		& ESBM & \cite{relin} & \cite{faces} & \cite{facese} \\
		\hline
		Overlap between top-5 summaries & 1.99 \tiny{(39.8$\%$)} & 2.91 \tiny{(58.2$\%$)} & 1.92 \tiny{(38.4$\%$)} & 2.12 \tiny{(42.4$\%$)} \\
		Overlap between top-10 summaries & 5.42 \tiny{(54.2$\%$)} & 7.86 \tiny{(78.6$\%$)} & 4.64 \tiny{(46.4$\%$)} & 5.44 \tiny{(54.4$\%$)} \\
		\hline
		Ground-truth summaries per entity & 6 & 4.43 & $\geq$ 7 & $\geq$ 4 \\
		\hline
	\end{tabular}
\end{table}

\section{Evaluating with ESBM}
\label{sec:evaluating}

We used ESBM to perform the most extensive evaluation of general-purpose entity summarizers to date. In this section, we will first describe evaluation criteria. Then we introduce the entity summarizers that we evaluate. Finally we present evaluation results.

\subsection{Evaluation Criteria}

Let~$S_m$ be a machine-generated entity summary. Let~$S_h$ be a human-made ground-truth summary. To compare~$S_m$ with~$S_h$ and assess the quality of~$S_m$ based on how much $S_m$~is close to~$S_h$, it is natural to compute precision~(P), recall~(R), and F1. The results are in the range of~0--1:
\begin{equation}
    \text{P} = \frac{|S_m \cap S_h|}{|S_m|} \,,\quad
    \text{R} = \frac{|S_m \cap S_h|}{|S_h|} \,,\quad
    \text{F1} = \frac{2 \cdot \text{P} \cdot \text{R}}{\text{P} + \text{R}} \,.
\end{equation}

In the experiments we configure entity summarizers to output at most $k$~triples and we set $k=|S_h|$, i.e.,~$k=5$ and $k=10$ are our two settings corresponding to the sizes of ground-truth summaries. We will trivially have P$=$R$=$F1 if $|S_m|=|S_h|$. However, some entity summarizers may output less than $k$~triples. For example, DIVERSUM~\cite{diversum} disallows an entity summary to contain triples having the same property. It is possible that an entity description contains less than $k$~distinct properties and hence DIVERSUM has to output less than $k$~triples. In this case, P$\neq$R and one should rely on F1.

In the evaluation, for each entity in ESBM, we compare a machine-generated summary with each of the 6~ground-truth summaries by calculating F1, and take their aggregation value. Finally we report the mean F1 over all the entities. For aggregation function, we report the results of average, to show an overall match with all the different ground truths; on the website we also give the results of maximum, to show the best match with each individual ground truth.

\subsection{Participating Entity Summarizers}

We not only evaluate existing entity summarizers but also compare them with two special entity summarizers we create: an oracle entity summarizer which is used to show the best possible performance on ESBM, and a new supervised learning based entity summarizer.

\textbf{Existing Entity Summarizers.}
We evaluate 9~out of the 12~general-purpose entity summarizers reviewed in Section~\ref{sec:related-work}. We re-implement RELIN~\cite{relin}, DIVERSUM~\cite{diversum}, LinkSUM~\cite{linksum}, FACES~\cite{faces}, FACES-E~\cite{facese}, and CD~\cite{cd}, while MPSUM~\cite{mpsum}, BAFREC~\cite{bafrec}, and KAFCA~\cite{kafca} are open source. We exclude SUMMARUM~\cite{summarum}, ES-LDA~\cite{eslda}, and ES-LDA$_{ext}$~\cite{esldaext} because LinkSUM represents an extension of SUMMARUM, and MPSUM represents an extension of ES-LDA and ES-LDA$_{ext}$.

We follow the original implementation and suggested configuration of existing entity summarizers as far as possible. However, for RELIN, we replace its Google-based relatedness measure with a string metric~\cite{isub} because Google's search API is no longer free. We also use this metric to replace the unavailable UMBC's SimService used in FACES-E. For DIVERSUM, we ignore its witness count measure since it does not apply to ESBM. For LinkSUM, we obtain backlinks between entities in LinkedMDB via their corresponding entities in DBpedia.

RELIN, CD, and LinkSUM compute a weighted combination of two scoring components. We tune these hyperparameters in the range of 0--1 in 0.01~increments. Since these summarizers are unsupervised, we use both the training set and the validation set described in Section~\ref{sec:creating-train-test} for tuning hyperparameters.

\textbf{Oracle Entity Summarizer.}
We implement an entity summarizer denoted by ORACLE to approximate the best possible performance on ESBM and form a reference point used for comparisons. ORACLE simply outputs $k$~triples that are selected by the most participants into ground-truth summaries.

\textbf{Supervised Learning Based Entity Summarizer.}
Existing general-purpose entity summarizers are unsupervised. We implement a supervised learning based entity summarizer with features that are used by existing entity summarizers. A triple with property~$p$ and value~$v$ describing entity~$e$ is represented by the following features:
\begin{itemize}
    \item $\globalfreq_\mathbb{T}$: the number of triples in the dataset where $p$~appears~\cite{linksum,bafrec},
    \item $\localfreq$: the number of triples in the description of~$e$ where $p$~appears~\cite{diversum,linksum},
    \item $\vfreq_\mathbb{T}$: the number of triples in the dataset where $v$~appears~\cite{faces,facese,bafrec}, and
    \item $\selfinfo$: the self-information of the triple~\cite{relin,faces,facese,cd}.
\end{itemize}
\noindent We also add three binary features:
\begin{itemize}
    \item $\isc$: whether $v$~is a class,
    \item $\ise$: whether $v$~is an entity, and
    \item $\isl$: whether $v$~is a literal.
\end{itemize}

Based on the training and validation sets described in Section~\ref{sec:creating-train-test}, we implement and tune 6~pointwise learning to rank models provided by Weka: SMOreg, LinearRegression, MultilayerPerceptron, AdditiveRegression, REPTree, and RandomForest. Each model outputs $k$~top-ranked triples as a summary.

\subsection{Evaluation Results}
We first report the overall evaluation results to show which entity summarizer generally performs better. Then we break down the results into different entity types (i.e.,~classes) for detailed comparison. Finally we present and analyze the performance of our supervised learning based entity summarizer.

\begin{table*}[!t]
	\caption{Average F1 over all the entities in a dataset. For the nine existing entity summarizers, significant improvements and losses over each other are indicated by~$\blacktriangle$ and~$\blacktriangledown$ ($p<0.05$), respectively. Insignificant differences are indicated by~$\circ$.}
	\label{tab:best}
	\centering
	\resizebox{\textwidth}{!}{
	\begin{tabular}{|c|c|c|c|c|}
		\hline
		& \multicolumn{2}{|c|}{DBpedia} & \multicolumn{2}{|c|}{LinkedMDB} \\
		\cline{2-5}
		& $k=5$ & $k=10$ & $k=5$ & $k=10$ \\
		\hline

RELIN & 0.242~\tiny{$^{\text{-}\circ\circ\blacktriangledown\blacktriangledown\blacktriangledown\blacktriangledown\blacktriangledown\blacktriangledown}$}		
 & 0.455~\tiny{$^{\text{-}\blacktriangledown\circ\circ\blacktriangledown\circ\blacktriangledown\blacktriangledown\blacktriangledown}$}		
 & 0.203~\tiny{$^{\text{-}\circ\circ\blacktriangledown\circ\blacktriangle\blacktriangledown\circ\blacktriangledown}$}		
 & 0.258~\tiny{$^{\text{-}\blacktriangledown\circ\blacktriangledown\blacktriangledown\circ\blacktriangledown\blacktriangledown\blacktriangledown}$}		
  \\			
DIVERSUM & 0.249~\tiny{$^{\circ\text{-}\circ\circ\blacktriangledown\blacktriangledown\blacktriangledown\blacktriangledown\blacktriangledown}$}		
 & 0.507~\tiny{$^{\blacktriangle\text{-}\blacktriangle\circ\circ\circ\circ\circ\circ}$}		
 & 0.207~\tiny{$^{\circ\text{-}\circ\blacktriangledown\circ\blacktriangle\blacktriangledown\circ\blacktriangledown}$}		
 & 0.358~\tiny{$^{\blacktriangle\text{-}\blacktriangle\circ\circ\blacktriangle\blacktriangledown\circ\blacktriangledown}$}		
  \\ 			
FACES & 0.270~\tiny{$^{\circ\circ\text{-}\circ\circ\circ\blacktriangledown\blacktriangledown\blacktriangledown}$}		
 & 0.428~\tiny{$^{\circ\blacktriangledown\text{-}\blacktriangledown\blacktriangledown\blacktriangledown\blacktriangledown\blacktriangledown\blacktriangledown}$}		
 & 0.169~\tiny{$^{\circ\circ\text{-}\blacktriangledown\blacktriangledown\circ\blacktriangledown\blacktriangledown\blacktriangledown}$}		
 & 0.263~\tiny{$^{\circ\blacktriangledown\text{-}\blacktriangledown\blacktriangledown\circ\blacktriangledown\blacktriangledown\blacktriangledown}$}		
  \\ 			
FACES-E & 0.280~\tiny{$^{\blacktriangle\circ\circ\text{-}\circ\circ\blacktriangledown\blacktriangledown\blacktriangledown}$}		
 & 0.488~\tiny{$^{\circ\circ\blacktriangle\text{-}\circ\circ\circ\circ\circ}$}		
 & 0.313~\tiny{$^{\blacktriangle\blacktriangle\blacktriangle\text{-}\blacktriangle\blacktriangle\blacktriangledown\blacktriangle\circ}$}		
 & 0.393~\tiny{$^{\blacktriangle\circ\blacktriangle\text{-}\blacktriangle\blacktriangle\circ\circ\circ}$}		
  \\ 		
  
CD & 0.283~\tiny{$^{\blacktriangle\blacktriangle\circ\circ\text{-}\circ\blacktriangledown\circ\circ}$}		
 & 0.513~\tiny{$^{\blacktriangle\circ\blacktriangle\circ\text{-}\circ\circ\circ\circ}$}		
 & 0.217~\tiny{$^{\circ\circ\blacktriangle\blacktriangledown\text{-}\blacktriangle\blacktriangledown\circ\blacktriangledown}$}		
 & 0.331~\tiny{$^{\blacktriangle\circ\blacktriangle\blacktriangledown\text{-}\blacktriangle\blacktriangledown\blacktriangledown\blacktriangledown}$}		
  \\			
LinkSUM & 0.287~\tiny{$^{\blacktriangle\blacktriangle\circ\circ\circ\text{-}\blacktriangledown\circ\circ}$}		
 & 0.486~\tiny{$^{\circ\circ\blacktriangle\circ\circ\text{-}\circ\circ\circ}$}		
 & 0.140~\tiny{$^{\blacktriangledown\blacktriangledown\circ\blacktriangledown\blacktriangledown\text{-}\blacktriangledown\blacktriangledown\blacktriangledown}$}		
 & 0.279~\tiny{$^{\circ\blacktriangledown\circ\blacktriangledown\blacktriangledown\text{-}\blacktriangledown\blacktriangledown\blacktriangledown}$}		
 \\			
BAFREC & 0.335~\tiny{$^{\blacktriangle\blacktriangle\blacktriangle\blacktriangle\blacktriangle\blacktriangle\text{-}\circ\circ}$}		
 & 0.503~\tiny{$^{\blacktriangle\circ\blacktriangle\circ\circ\circ\text{-}\circ\circ}$}		
 & 0.360~\tiny{$^{\blacktriangle\blacktriangle\blacktriangle\blacktriangle\blacktriangle\blacktriangle\text{-}\blacktriangle\blacktriangle}$}		
 & 0.402~\tiny{$^{\blacktriangle\blacktriangle\blacktriangle\circ\blacktriangle\blacktriangle\text{-}\circ\circ}$}		
  \\ 			
KAFCA & 0.314~\tiny{$^{\blacktriangle\blacktriangle\blacktriangle\blacktriangle\circ\circ\circ\text{-}\circ}$}		
 & 0.509~\tiny{$^{\blacktriangle\circ\blacktriangle\circ\circ\circ\circ\text{-}\circ}$}		
 & 0.244~\tiny{$^{\circ\circ\blacktriangle\blacktriangledown\circ\blacktriangle\blacktriangledown\text{-}\circ}$}		
 & 0.397~\tiny{$^{\blacktriangle\circ\blacktriangle\circ\blacktriangle\blacktriangle\circ\text{-}\circ}$}		
  \\ 			
MPSUM & 0.314~\tiny{$^{\blacktriangle\blacktriangle\blacktriangle\blacktriangle\circ\circ\circ\circ\text{-}}$}		
 & 0.512~\tiny{$^{\blacktriangle\circ\blacktriangle\circ\circ\circ\circ\circ\text{-}}$}		
 & 0.272~\tiny{$^{\blacktriangle\blacktriangle\blacktriangle\circ\blacktriangle\blacktriangle\blacktriangledown\circ\text{-}}$}		
 & 0.423~\tiny{$^{\blacktriangle\blacktriangle\blacktriangle\circ\blacktriangle\blacktriangle\circ\circ\text{-}}$}		
  \\ 					

\hline
ORACLE & 0.595 & 0.713 & 0.619 & 0.678  \\ 
\hline
SMOreg & 0.279 & 0.543 & 0.403 & 0.472 \\ 
LinearRegression & 0.319 & 0.556 & 0.401 & 0.471 \\ 
MultilayerPerceptron & 0.340 & 0.560 & 0.390 & 0.477 \\ 
AdditiveRegression & 0.345 & 0.558 & 0.415 & 0.510 \\ 
REPTree & 0.392 & 0.570 & 0.455 & 0.538 \\ 
RandomForest & 0.399 & 0.576 & 0.449 & 0.506 \\ 
		\hline		
	\end{tabular}
	}
\end{table*}

\textbf{Overall Results of Existing Entity Summarizers.}
Table~\ref{tab:best} presents the results of all the participating entity summarizers on two datasets under two size constraints. We compare nine existing summarizers using one-way ANOVA post-hoc LSD and we show whether the difference between each pair of them is statistical significant at the 0.05~level. Among existing summarizers, BAFREC achieves the highest F1 under $k=5$. It significantly outperforms six existing summarizers on DBpedia and outperforms all the eight ones on LinkedMDB. It is also among the best under $k=10$. MPSUM follows BAFREC under $k=5$ but performs slightly better under $k=10$. Other top-tier results belong to KAFCA on DBpedia and FACES-E on LinkedMDB.

The F1 scores of ORACLE are in the range of~0.595--0.713. It is impossible for ORACLE or any other summarizer to reach $\text{F1}=1$, because for each entity in ESBM there are six ground-truth summaries which are often different and hence cannot simultaneously match a machine-generated summary. However, the gap between the results of ORACLE and the best results of existing summarizers is still as large as~0.20--0.26, suggesting that there is much room for improvement.

\begin{figure}[!t]
\centering
\includegraphics[width=0.95\linewidth]{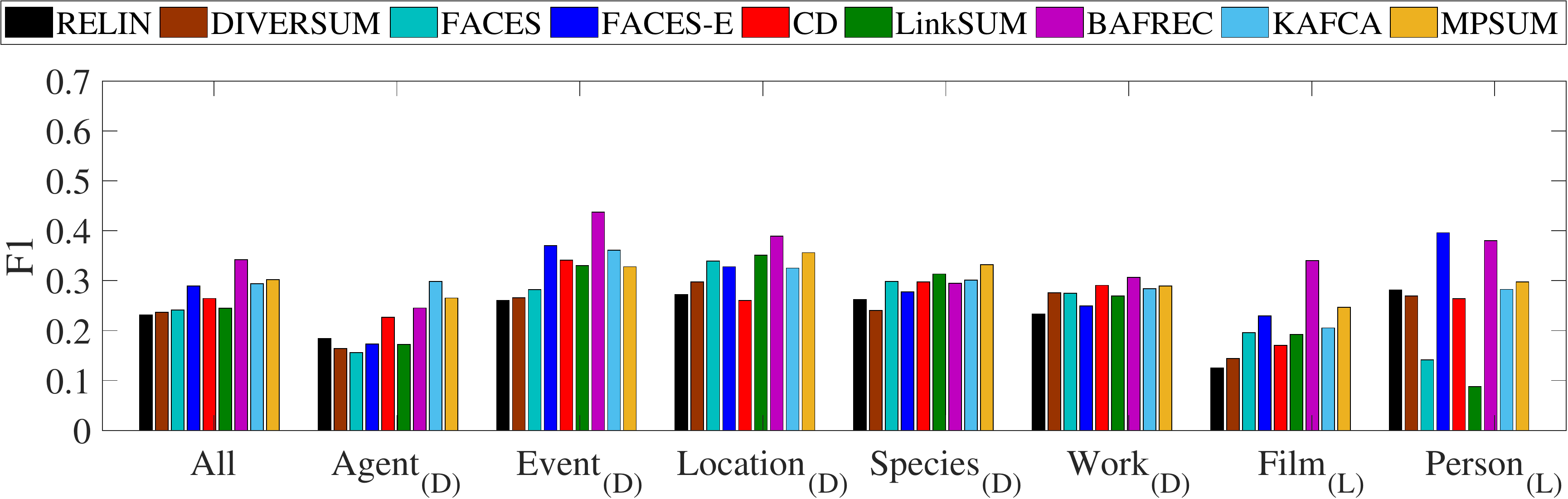}
\caption{Average F1 over all the entities in each class under $k=5$.}
\label{fig:set_typeF_top5}
\end{figure}

\begin{figure}[!t]
\centering
\includegraphics[width=0.95\linewidth]{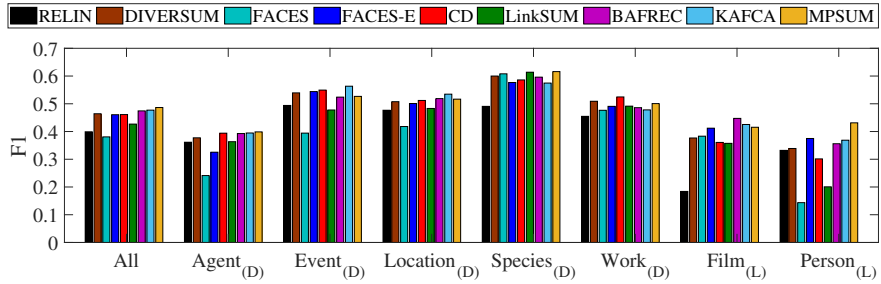}
\caption{Average F1 over all the entities in each class under $k=10$.}
\label{fig:set_typeF_top10}
\end{figure}

\textbf{Results on Different Entity Types.}
We break down the results of existing entity summarizers into 7~entity types (i.e.,~classes). When $k=5$ in Fig.~\ref{fig:set_typeF_top5}, there is no single winner on every class, but BAFREC and MPSUM are among top three on 6~classes, showing relatively good generalizability over different entity types. Some entity summarizers have limited generalizability and they perform not well on certain classes. For example, RELIN and CD mainly rely on the self-information of a triple, while for \texttt{Location} entities their latitudes and longitudes are often unique in DBpedia but such triples with large self-information rarely appear in ground-truth summaries. Besides, most summarizers generate low-quality summaries for \texttt{Agent}, \texttt{Film}, and \texttt{Person} entities. This is not surprising since these entities are described in more triples and/or by more properties according to Fig.~\ref{fig:composition}. Their summarization is inherently more difficult. When $k=10$ in Fig.~\ref{fig:set_typeF_top10}, MPSUM is still among top three on 6~classes. KAFCA also shows relatively good generalizability---among top three on 5~classes.


\textbf{Results of Supervised Learning.}
As shown in Table~\ref{tab:best}, among the six supervised learning based methods, RandomForest and REPTree achieve the highest F1 on DBpedia and LinkedMDB, respectively. Four methods (MultilayerPerceptron, AdditiveRegression, REPTree, and RandomForest) outperform all the existing entity summarizers on both datasets under both size constraints, and two methods (SMOreg and LinearRegression) only fail to outperform in one setting. The results demonstrate the powerfulness of supervised learning for entity summarization. Further, recall that these methods only use standard models and rely on features that are used by existing entity summarizers. It would be reasonable to predict that better results can be achieved with specialized models and more advanced features. However, creating a large number of ground-truth summaries for training is expensive, and the generalizability of supervised methods for entity summarization still needs further exploration.

Moreover, we are interested in how much the seven features contribute to the good performance of supervised learning. Table~\ref{table:rmf} shows the results of RandomForest after removing each individual feature. Considering statistical significance at the 0.05~level, two features~$\globalfreq_\mathbb{T}$ and~$\localfreq$ show effectiveness on both datasets under both size constraints, and two features~$\vfreq_\mathbb{T}$ and~$\selfinfo$ are only effective on LinkedMDB. The usefulness of the three binary features~$\isc$, $\ise$, and~$\isl$ is not statistically significant.

\begin{table}[!t] 
	\caption{F1 of RandomForest after removing each individual feature, its difference from using all features~($\Delta\%$), and the significance level for the difference~($p$).}
	\label{table:rmf}
	\centering
    \resizebox{\textwidth}{!}{
	\begin{tabular}{|l|rrr|l|rrr|l|rrr|l|rrr|}
		\hline
		\multicolumn{8}{|c|}{DBpedia} & \multicolumn{8}{|c|}{LinkedMDB} \\
		\hline
		\multicolumn{4}{|c|}{$k=5$} & \multicolumn{4}{|c|}{$k=10$} & \multicolumn{4}{|c|}{$k=5$} & \multicolumn{4}{|c|}{$k=10$} 	\\
		\hline
 & \texttt{F1} & $\Delta\%$ & $p$ &  & \texttt{F1} & $\Delta\%$ & $p$ &  & \texttt{F1} & $\Delta\%$ & $p$ &  & \texttt{F1} & $\Delta\%$ & $p$  \\
 \hline
 All & 0.399 & --- & --- & All & 0.576 & --- & --- & All & 0.449 & --- & --- & All & 0.506 & --- & --- \\
\hline
-$\globalfreq_\mathbb{T}$ & 0.346  & $-$5.360  & 0.000  & -$\localfreq$ & 0.546  & $-$0.030  & 0.000  & -$\globalfreq_\mathbb{T}$ & 0.383  & $-$0.066  & 0.000  & -$\localfreq$ & 0.473  & $-$0.033  & 0.008 \\ 
-$\localfreq$ & 0.366  & $-$3.307  & 0.000  & -$\globalfreq_\mathbb{T}$ & 0.551  & $-$0.025  & 0.000  & -$\localfreq$ & 0.413  & $-$0.036  & 0.025  & -$\vfreq_\mathbb{T}$ & 0.477  & $-$0.029  & 0.010 \\ 
-$\isc$ & 0.392  & $-$0.720  & 0.261  & -$\vfreq_\mathbb{T}$ & 0.569  & $-$0.007  & 0.198  & -$\vfreq_\mathbb{T}$ & 0.414  & $-$0.035  & 0.022  & -$\globalfreq_\mathbb{T}$ & 0.479  & $-$0.027  & 0.007 \\ 
-$\ise$ & 0.397  & $-$0.267  & 0.720  & -$\ise$ & 0.570  & $-$0.006  & 0.262  & -$\selfinfo$ & 0.442  & $-$0.007  & 0.574  & -$\selfinfo$ & 0.486  & $-$0.020  & 0.009 \\ 
-$\selfinfo$ & 0.400  & $+$0.027  & 0.973  & -$\isc$ & 0.571  & $-$0.005  & 0.303  & -$\ise$ & 0.455  & $+$0.005  & 0.651  & -$\isl$ & 0.491  & $-$0.015  & 0.079 \\ 
-$\isl$ & 0.401  & $+$0.160  & 0.816  & -$\selfinfo$ & 0.572  & $-$0.004  & 0.402  & -$\isl$ & 0.456  & $+$0.007  & 0.504  & -$\ise$ & 0.492  & $-$0.014  & 0.148 \\ 
-$\vfreq_\mathbb{T}$ & 0.407  & $+$0.720  & 0.346  & -$\isl$ & 0.578  & $+$0.002  & 0.683  & -$\isc$ & 0.463  & $+$0.013  & 0.281  & -$\isc$ & 0.514  & $+$0.008  & 0.396 \\ 
\hline
	\end{tabular}
	}
\end{table}

\textbf{Conclusion.}
Among existing entity summarizers, BAFREC generally shows the best performance on ESBM while MPSUM seems more robust. However, none of them are comparable with our straightforward implementation of supervised learning, which in turn is still far away from the best possible performance represented by ORACLE. Therefore, entity summarization on ESBM is a non-trivial task. We invite researchers to experiment with new ideas on ESBM.
\section{Discussion and Future work}
\label{sec:discussion-future-work}

We identify the following limitations of our work to be addressed in future work.

\textbf{Evaluation Criteria.}
We compute F1 score in the evaluation, which is based on common triples but ignores semantic overlap between triples. A triple~$t$ in a machine-generated summary~$S$ may partially cover the information provided by some triple~$t'$ in the ground-truth summary. It may be reasonable to not completely penalize~$S$ for missing~$t'$ but give some reward for the presence of~$t$. However, it is difficult to quantify the extent of penalization for all possible cases, particularly when multiple triples semantically overlap with each other. In future work, we will explore more proper evaluation criteria.

\textbf{Representativeness of Ground Truth.}
The ground-truth summaries in ESBM are not supposed to represent the view of the entire user population.
They are intrinsically biased towards their creators.
Besides, these ground-truth summaries are created for general purposes. Accordingly, we use them to evaluate general-purpose entity summarizers. However, for a specific task, these summaries may not show optimality, and the participating systems may not represent the state of the art. Still, we believe it is valuable to evaluate general-purpose systems not only because of their wide range of applications but also because their original technical features have been reused by task-specific systems. In future work, we will extend ESBM to a larger scale, and will consider benchmarking task-specific entity summarization.

\textbf{Form of Ground Truth.}
ESBM provides ground-truth summaries, whereas some other benchmarks offer ground-truth scores of triples~\cite{game,assign,franco}. Scoring-based ground truth may more comprehensively evaluate an entity summarizer than our set-based ground truth because it not only considers the triples in a machine-generated summary but also assesses the rest of the triples. However, on the other hand, a set of top-scored triples may not equal an optimal summary because they may cover limited aspects of an entity and show redundancy. Therefore, both methods have their advantages and disadvantages. In future work, we will conduct scoring-based evaluation to compare with the current results.

\section*{Acknowledgments}
This work was supported in part by the NSFC under Grant 61772264 and in part by the Qing Lan Program of Jiangsu Province.

\bibliographystyle{splncs04.bst}
\bibliography{main}

\begin{thebibliography}{10}
\providecommand{\url}[1]{\texttt{#1}}
\providecommand{\urlprefix}{URL }
\providecommand{\doi}[1]{https://doi.org/#1}

\bibitem{franco}
Bobic, T., Waitelonis, J., Sack, H.: {FRanCo} - {A} ground truth corpus for
  fact ranking evaluation. In: {SumPre} 2015 \& {HSWI} 2015 (2015)

\bibitem{relin}
Cheng, G., Tran, T., Qu, Y.: {RELIN:} relatedness and informativeness-based
  centrality for entity summarization. In: {ISWC} 2011, Part {I}. pp. 114--129
  (2011). \doi{10.1007/978-3-642-25073-6\_8}

\bibitem{entityresolution}
Cheng, G., Xu, D., Qu, Y.: {C3D+P:} {A} summarization method for interactive
  entity resolution. J. Web Sem.  \textbf{35},  203--213 (2015).
  \doi{10.1016/j.websem.2015.05.004}

\bibitem{entitylinking}
Cheng, G., Xu, D., Qu, Y.: Summarizing entity descriptions for effective and
  efficient human-centered entity linking. In: {WWW} 2015. pp. 184--194 (2015).
  \doi{10.1145/2736277.2741094}

\bibitem{timeline}
Gottschalk, S., Demidova, E.: {EventKG} - the hub of event knowledge on the web
  - and biographical timeline generation. Semantic Web  \textbf{10}(6),
  1039--1070 (2019). \doi{10.3233/SW-190355}

\bibitem{thesiskalpa}
Gunaratna, K.: Semantics-based Summarization of Entities in Knowledge Graphs.
  Ph.D. thesis, Wright State University (2017)

\bibitem{faces}
Gunaratna, K., Thirunarayan, K., Sheth, A.P.: {FACES:} diversity-aware entity
  summarization using incremental hierarchical conceptual clustering. In:
  {AAAI} 2015. pp. 116--122 (2015)

\bibitem{facese}
Gunaratna, K., Thirunarayan, K., Sheth, A.P., Cheng, G.: Gleaning types for
  literals in {RDF} triples with application to entity summarization. In:
  {ESWC} 2016. pp. 85--100 (2016). \doi{10.1007/978-3-319-34129-3\_6}

\bibitem{dynes}
Hasibi, F., Balog, K., Bratsberg, S.E.: Dynamic factual summaries for entity
  cards. In: {SIGIR} 2017. pp. 773--782 (2017). \doi{10.1145/3077136.3080810}

\bibitem{lmdb}
Hassanzadeh, O., Consens, M.P.: Linked movie data base. In: {LDOW} 2009 (2009)

\bibitem{kafca}
Kim, E.K., Choi, K.S.: Entity summarization based on formal concept analysis.
  In: {EYRE} 2018 (2018)

\bibitem{bafrec}
Kroll, H., Nagel, D., Balke, W.T.: {BAFREC}: Balancing frequency and rarity for
  entity characterization in linked open data. In: {EYRE} 2018 (2018)

\bibitem{assign}
Langer, P., Schulze, P., George, S., Kohnen, M., Metzke, T., Abedjan, Z.,
  Kasneci, G.: Assigning global relevance scores to {DBpedia} facts. In: {ICDE}
  Workshops 2014. pp. 248--253 (2014). \doi{10.1109/ICDEW.2014.6818334}

\bibitem{dbpedia}
Lehmann, J., Isele, R., Jakob, M., Jentzsch, A., Kontokostas, D., Mendes, P.N.,
  Hellmann, S., Morsey, M., van Kleef, P., Auer, S., Bizer, C.: {DBpedia} - {A}
  large-scale, multilingual knowledge base extracted from {Wikipedia}. Semantic
  Web  \textbf{6}(2),  167--195 (2015). \doi{10.3233/SW-140134}

\bibitem{survey}
Liu, Q., Cheng, G., Gunaratna, K., Qu, Y.: Entity summarization: State of the
  art and future challenges. CoRR  \textbf{abs/1910.08252} (2019),
  \url{http://arxiv.org/abs/1910.08252}

\bibitem{esldaext}
Pouriyeh, S.A., Allahyari, M., Kochut, K., Cheng, G., Arabnia, H.R.: Combining
  word embedding and knowledge-based topic modeling for entity summarization.
  In: {ICSC} 2018. pp. 252--255 (2018). \doi{10.1109/ICSC.2018.00044}

\bibitem{eslda}
Pouriyeh, S.A., Allahyari, M., Kochut, K., Cheng, G., Arabnia, H.R.: {ES-LDA:}
  entity summarization using knowledge-based topic modeling. In: {IJCNLP} 2017,
  Volume 1. pp. 316--325 (2017)

\bibitem{benchmark}
Sim, S.E., Easterbrook, S.M., Holt, R.C.: Using benchmarking to advance
  research: {A} challenge to software engineering. In: {ICSE} 2003. pp. 74--83
  (2003). \doi{10.1109/ICSE.2003.1201189}

\bibitem{isub}
Stoilos, G., Stamou, G.B., Kollias, S.D.: A string metric for ontology
  alignment. In: {ISWC} 2005. pp. 624--637 (2005). \doi{10.1007/11574620\_45}

\bibitem{diversum}
Sydow, M., Pikula, M., Schenkel, R.: The notion of diversity in graphical
  entity summarisation on semantic knowledge graphs. J. Intell. Inf. Syst.
  \textbf{41}(2),  109--149 (2013). \doi{10.1007/s10844-013-0239-6}

\bibitem{thesisandreas}
Thalhammer, A.: Linked Data Entity Summarization. Ph.D. thesis, Karlsruher
  Institut f\"ur Technologie (2017)

\bibitem{game}
Thalhammer, A., Knuth, M., Sack, H.: Evaluating entity summarization using a
  game-based ground truth. In: {ISWC} 2012, Part {II}. pp. 350--361 (2012).
  \doi{10.1007/978-3-642-35173-0\_24}

\bibitem{linksum}
Thalhammer, A., Lasierra, N., Rettinger, A.: {LinkSUM}: Using link analysis to
  summarize entity data. In: {ICWE} 2016. pp. 244--261 (2016).
  \doi{10.1007/978-3-319-38791-8\_14}

\bibitem{summarum}
Thalhammer, A., Rettinger, A.: Browsing {DBpedia} entities with summaries. In:
  {ESWC} 2014 Satellite Events. pp. 511--515 (2014).
  \doi{10.1007/978-3-319-11955-7\_76}

\bibitem{movie}
Thalhammer, A., Toma, I., Roa{-}Valverde, A.J., Fensel, D.: Leveraging usage
  data for linked data movie entity summarization. In: {USEWOD} 2012 (2012)

\bibitem{trank}
Tonon, A., Catasta, M., Prokofyev, R., Demartini, G., Aberer, K.,
  Cudr{\'{e}}{-}Mauroux, P.: Contextualized ranking of entity types based on
  knowledge graphs. J. Web Sem.  \textbf{37-38},  170--183 (2016).
  \doi{10.1016/j.websem.2015.12.005}

\bibitem{mpsum}
Wei, D., Gao, S., Liu, Y., Liu, Z., Huang, L.: {MPSUM}: Entity summarization
  with predicate-based matching. In: {EYRE} 2018 (2018)

\bibitem{cd}
Xu, D., Zheng, L., Qu, Y.: {CD} at {ENSEC} 2016: Generating characteristic and
  diverse entity summaries. In: {SumPre} 2016 (2016)

\end{thebibliography}

\end{document}